\documentstyle[12pt]{article}

\begin{document}

\setcounter{page}{1}

\centerline{\large {\bf How to Measure the
$Q\overline{Q}$ Potential in a Light-Front Calculation}\hspace{-.2cm}
\footnote{Talk presented at ``Theory of Hadrons and Light-Front QCD'',
Polona Zgorzelisko, Poland, Aug. 94.}}
\vskip.3in
\centerline{ Matthias Burkardt}
\centerline{ Institute for Nuclear Theory,
University of Washington, NK-12, Seattle, WA 98195}

\vskip.5 in
\centerline{\bf Abstract}
\vskip.1in
A scheme is developed which shows how one would, given a light-front
Hamiltonian for
QCD, extract the $Q\overline{Q}$
potential, i.e. the quantity which corresponds to the potential
between two infinitely heavy quarks in a rest frame, from a
light-front calculation.
The resulting potential will in general
depend on the direction along which the infinitely
heavy sources are separated and thus provides a direct probe of
violations of rotational invariance in a physical observable.
Furthermore, easy comparison to data from $c\overline{c}$
and $b\overline{b}$ spectroscopy and to lattice data is possible.
The scheme may thus be very helpful in constructing a light-front
Hamiltonian through an iterative procedure.
\vskip.1in

\vskip.3in
{\bf 1. Introduction}
\vskip.1in

\noindent
The central quantity to be discussed in these lectures is:
\vskip.1in
$V(\vec{R}) \equiv$ 
potential energy
of an $\infty$-heavy $Q\overline{Q}$-pair at separation
${\vec R}$ in its rest frame.
\vskip.1in
\noindent
There are several reasons to study this observable in the
light-front (LF)
framework
\begin{itemize}
\item The LF formalism lacks manifest rotational invariance.
Therefore, if one starts with a {\it wrong} LF Hamiltonian
for QCD, the result $V(\vec{R})$ depends on the orientation
of ${\vec R}$ with respect to the 3-axis. \footnote{We use the
notation $A^\pm = A_\mp = (A^0\pm A^3)/\sqrt{2}$,
${\vec A}_\perp = (A^1,A^2)$.}
A measurement of $V(\vec{R})$ thus provides a direct probe of
rotational invariance in a physical observable.
\item Calculating $V(\vec{R})$ on the LF allows one to test
both short (asymptotic freedom) and long distance (confinement)
aspects of the theory within the same set of calculations by
simply changing $|{\vec R}|,$ and thus provides a stringent
test for perturbative as well as nonperturbative aspects of the
interactions.
\item  And most importantly, $V(\vec{R})$ in QCD is very well known over
a large range of distances from the spectroscopy of heavy
$Q\overline{Q}$ mesons as well as from nonperturbative Euclidean
lattice calculations (at least in the absence of dynamical quarks
--- but it is easy to make the same approximation in a LF framework).
\end{itemize}
\newpage
\unitlength1.cm
\begin{figure}
\begin{picture}(10.,5.4)(-5,-1.)
\put(4.,.5){\vector(1,0){7.}}
\put(7.5,.5){\vector(0,1){4.}}
\put(7.5,.5){\vector(1,1){3.5}}
\put(7.5,.5){\vector(-1,1){3.5}}
\put(7.8,3.5){\makebox(0,0){$x^0$}}
\put(10.5,.1){\makebox(0,0){$x^3$}}
\put(10.,1.8){\makebox(0,0){$x^+\equiv\frac{x^0+x^3}{\sqrt{2}}$}}
\put(4.2,2.5){\makebox(0,0){$x^-\equiv\frac{x^0-x^3}{\sqrt{2}}$}}
\thicklines \put(5.5,.5){\line(0,1){4.}}
\thicklines \put(7.4,.5){\line(0,1){4.}}
\put(6.,0){\vector(-1,0){.5}}
\put(7.,0){\vector(1,0){.4}}
\put(6.6,.1){\makebox(0,0){$\Delta x^3$}}
\put(6.2,2.3){\vector(-1,1){.7}}
\put(6.8,1.7){\vector(1,-1){.6}}
\put(6.7,2.1){\makebox(0,0){$\Delta x^-$}}
\put(7.2,-.8){\makebox(0,0){\parbox[t]{7.9cm}{Fig.1: World lines for
two charges with longitudinal
separation $\Delta x^3$ in the rest frame.}}}
\put(-1.2,1.55){\makebox(0,0){\parbox{7.65cm}{
Before we embark on deriving the effective LF Hamiltonian
for two infinitely heavy sources, it is instructive to
understand physically what it means to have two fixed
sources at rest from the LF point-of-view.
As should be clear from Fig.1, fixed charges in a rest frame
correspond to charges that move with constant velocity
on the LF [$v^+=v^-=1/\sqrt{2}$ in the example in
Fig.1].
Furthermore, if the longitudinal separation was $\Delta x^3$ in
the rest frame, the charges have fixed separation
$\Delta x^-=\sqrt{2}\Delta x^3$ in the
}}}
\end{picture}
\label{world:fig}
\end{figure}
\noindent
longitudinal LF direction
[in the more general case, where the charges were moving
with constant four velocity $v^\mu$, where ${\vec v}_\perp=0$
in the rest frame, one obtains $\Delta x^- = \Delta x^3/v^+$].

The transverse
separation is the same on the LF as it is in a rest frame
description.
Therefore, in order to understand the physics of fixed charges
with separation ${\vec R}=(R^1,R^2,R^3)$ in a LF description,
we must first understand how to describe a ``dumbbell,''
with ends separated by $(\Delta x^1,\Delta x^2,\Delta x^-)
=(R^1,R^2,R^3/v^+)$,
that moves
with constant velocity $v^+$.

\vskip.3in
{\bf 2. One Heavy Quark on the LF}
\vskip.1in

A pair of fixed sources can also be interpreted (and treated)
as one {\it extended} source moving with constant velocity $v^\mu$
(for simplicity, we will keep ${\vec v}_\perp ={\vec 0}$).
This is reminiscent of heavy quarks and thus, as a warmup
exercise, it is very instructive to consider one (pointlike) heavy
quark on the LF first (see also Refs.[1,2]).

For simplicity, we will first take the heavy quark limit
for the canonical Hamiltonian, which can be written in the form
\begin{equation}
P^-_B = \frac{M_b^2+{\vec k}_{b\perp}^2}{2p_b^+}
+P^-_{HL}+P^-_{LL},
\label{eq:pminus}
\end{equation}
where $B$ represents the hadron, $b$ is the heavy quark,
$P^-_{HL}$ contains the interactions between heavy ($b$) and
light degrees of freedom and $P^-_{LL}$ contains all
terms involving light degrees of freedom only.
The heavy quark limit is obtained by making an expansion in
inverse powers of the $b$-quark mass. For this purpose we write
\begin{equation}
p^+_b=P^+_B - p^+_L=M_Bv^+-p_L^+,
\label{eq:P+}
\end{equation}
where $p_L^+$ is defined to be the sum of the longitudinal
(LF-) momenta of all light degrees of freedom. For the (total)
LF-energy we write on the l.h.s. of Eq.(\ref{eq:pminus})
\begin{equation}
P^-_B=M_Bv^-=\frac{M_B}{2v^+}= \frac{M_b+\delta E}{2v^+},
\label{eq:P-}
\end{equation}
where $\delta E \equiv M_B-M_b$ is the ``binding energy'' of
the hadron.
After inserting Eqs.(\ref{eq:P+}) and (\ref{eq:P-}) into Eq.(\ref{eq:pminus})
and expanding one obtains
\begin{eqnarray}
\frac{M_b+\delta E}{2v^+}
&=&
\frac{M_b^2 + {\vec k}_{b\perp}^2}{2(M_Bv^+-p_L^+)} +P^-_{HL} + P^-_{LL}
=\frac{M_b^2 + {\vec k}_{b\perp}^2}{2\left(M_bv^++\delta E v^+-p_L^+\right)}
+P^-_{HL} + P^-_{LL}
\nonumber\\
&=&
\frac{M_b}{2v^+} - \frac{\delta E}{2v^+}+ \frac{p_L^+}{2v^{+2}}
+{\cal O}(1/M_b)+P^-_{HL} + P^-_{LL}.
\label{eq:expand}
\end{eqnarray}
Note that we assumed that the transverse momentum of the
heavy quark is small compared to its mass which is justified
in a frame where the transverse velocity of the heavy
hadron vanishes.
The term proportional to $M_b$ cancels between the l.h.s.
and the r.h.s. of Eq.(\ref{eq:expand}) and we are left with
\begin{equation}
\frac{\delta E}{v^+}= \frac{p_L^+}{2v^{+2}}
+{\cal O}(1/M_b)+P^-_{HL} + P^-_{LL}.
\label{eq:deltae}
\end{equation}
The {\it brown muck} Hamiltonian $P^-_{LL}$ is the same as
for light-light systems and will not be discussed here.
The interaction term between the heavy quark and the
brown muck ($P^-_{HL}$) is more tedious but straightforward.
For example, heavy quark pair creation terms (via instantaneous
gluons) are proportional to $1/(p^+_{b_1}+p^+_{b_2})^2
\propto 1/M_b^2$ and can thus be neglected. Similarly, pair
creation of heavy quarks from virtual gluons is also suppressed
by at least one power of $M_b$. This also justifies our omission
of states containing more than one heavy quark from the
start.
Other terms that vanish
in $P^-_{HL}$ include interactions that involve
{\it instantaneous exchanges} of heavy quarks, which are
typically proportional to the inverse $p^+$ of the
exchanged quark and thus of the order ${\cal O}(M_b^{-1})$.
Up to this point, all interaction terms that we have
considered vanish in the heavy quark limit. The more interesting
ones are of course those terms which survive.
The simplest ones are the instantaneous gluon exchange
interactions with light quarks or gluons, which are,
respectively,
$V_{Qq}\propto (p_q^+-p_q^{'+})^{-2}$ and
$V_{Qg}\propto (p^+_g+p'^+_g){(p_g^+-p'^+_g)}^{-2}$ and
remain unchanged in the limit $M_b\rightarrow \infty$.
Terms which involve instantaneous gluon exchange and are
off-diagonal in the brown muck Fock space behave in the same
way.

The quark gluon vertex simplifies considerably. For finite
quark mass one has for the matrix element for the emission of
a gluon with momentum $k$, polarization $i$ and color $a$ between quarks
of momentum $p_1$ and $p_2$
\begin{equation}
P^-_{QQg} = -igT^a \left\{
2\frac{k^i}{k^+} -
\frac{ {\vec \sigma}_\perp {\vec p}_{2\perp}-iM_b}{p_2^+}
\sigma^i  -\sigma^i \frac{ {\vec \sigma}_\perp {\vec p}_{1\perp}+iM_b}{p_1^+}
\right\}
,
\end{equation}
where spinors as well as creation/destruction operators have been
omitted for simplicity. In the heavy quark limit [note
$1/p_1^+ -1/p_2^+ = {\cal O}(M_b^{-2})$] the spin dependent terms
drop out and one finds
\begin{equation}
P^-_{qqg} = -igT^a 2\frac{k^i}{k^+} .
\end{equation}
The spin of the heavy quark thus decouples completely, giving rise to
the well known $SU(2N_f)$ symmetry in heavy quark systems.

\vskip.3in
{\bf 3. Divergences}
\vskip.1in

Throughout the above discussion we have tacitly assumed that
all momentum scales other than the heavy quark longitudinal momentum
are small compared to the heavy quark mass (or longitudinal momentum).
To justify this assumption we impose the following cutoffs:
\begin{itemize}
\item a small momentum cutoff $\Theta( k_i^+ - \varepsilon v^+)$ on all
light constituents and on all momenta exchanged in instantaneous interactions.
\item a cutoff on the maximum allowed difference in kinetic energy at
each vertex\\
$\Theta\left( \frac{\Lambda}{v^+} - \left| P^-_{kin}(in)-P^-_{kin}(out)
\right| \right)$
\end{itemize}
Notice that the cutoffs have been chosen such that the spectrum (with
the regulators present) is manifestly independent of $v^+$.
Notice further that the above cutoffs guarantee that the heavy quark
limit commutes with loop integrations. In other words, for loop integrals
it does not matter whether one first computes the amplitudes
using the full Hamiltonian and then takes the heavy quark limit or whether one
takes the heavy quark limit first and computes loops using the
heavy quark effective LF Hamiltonian. This guarantees that
the heavy quark limit is nonsingular.

It should be emphasized that, while the above conditions are sufficient to
guarantee
a meaningful heavy quark limit, other, less stringent, ways to cutoff
the divergences (e.g. using a transverse lattice in combination with
some low-$k^+$ regulator) are conceivable.

In Section 2., we illustrated the heavy quark limit starting from
the {\it canonical} LF Hamiltonian. However, most regularization schemes,
particularly momentum cutoffs such as the one described above,
require non-canonical counterterms, including ``counterterm functions'' [4].
How does this influence the heavy quark limit? The counterterm functions
typically depend on ratios of momenta and one might be worried that
this provides some ``hidden'' dependence on the heavy quark mass,
because heavy quark momenta are of the order of the heavy quark mass.
It turns out that, with the above or similar cutoffs, heavy quark
momenta typically enter the counterterm functions
through ratios between two heavy quark momenta, but not ratios of
heavy quark to brown muck momenta.
\footnote{I was not able to give a strict proof for this intuitively obvious
result but could not find a counterexample either.} Upon expanding the
heavy quark momentum in powers of $1/M_b$ one
thus finds that $\log$'s of $M_b$ cancel in the effective LF Hamiltonian and
only ${\cal O}(1/M_b)$ corrections remain from the dependence of the
counterterm functions on the heavy quark momentum. This is just another
example of the fact that, with the above cutoffs in place, the heavy quark
limit and loop integrations commute.  Therefore,
as long as one keeps  $\Lambda \ll M_b$:
\\[1.5ex]
\fbox{\parbox[t]{15.5cm}{
the conterterm functions, which arise from renormalizing
the LF-Hamiltonian using noncovariant cutoffs, depend,
to leading order in $1/M_b$, only on ratios of brown
muck momenta but not on the heavy quark momentum or its mass.
}}\\[1.5ex]

By restricting all momenta (except the longitudinal momentum of the
heavy quark) and momentum transfers to the heavy quark to be small compared
to the heavy quark mass, we have implicitly limited the applications of
the above formalism to physical processes which satisfy these conditions.
In particular, we cannot use the above formalism directly for calculating
Isgur-Wise functions or decay constants for heavy mesons because these
processes typically involve a large momentum transfer.
The formalism is applicable, however, for calculating $B$-meson spectra as well
as the $Q\bar{Q}$-potential at distances that are
large compared to the inverse heavy quark mass.

For finite heavy quark masses, the total momentum that can be
carried by the brown muck is bounded from above by the total
momentum of the hadron. Since the total momentum of the hadron in
the heavy quark limit is infinite, this natural cutoff no longer
exists. However, this does {\it not} lead to any new divergences:
One can solve analytically the problem of photons coupled to
an infinitely heavy electron, where one obtains a photon distribution
that falls of at large $k^+$ like
$\rho(k^+) \equiv \int d k_\perp^2 \rho(k^+,k_\perp^2)
\sim 1/\left(v^+\Lambda {k^+}^4\right) + 1/{k^+}^5\log \left(\Lambda
v^+/k^+\right)$.
In QCD, if one excludes the strict
chiral limit, there are no massless excitations and one expects
an even more rapid falloff at large $k^+$.

\vskip.3in
{\bf 4. Two Heavy Sources}
\vskip.1in

As we discussed in section 2, two heavy sources at fixed
separation can be formally treated as one extended heavy source
\footnote{Just think of a dumbbell.}.
Therefore, the LF Hamiltonian for two sources is the same as for one
source with two minor modifications:
\begin{itemize}
\item All vertices involving a heavy source get modified
according to the rule
\begin{eqnarray}
\left\{
{\chi^\dagger}' T^a \chi \,\times {\hat O}( brown\, muck)
\right\} + h.c.\rightarrow \quad\quad \quad\quad \quad\quad
\quad\quad\quad\quad \quad\quad \quad\quad
\nonumber\\
\left\{
\left[ {\chi_Q^\dagger}' T^a \chi_Q F_R(q)
-\chi_{\bar{Q}}^\dagger T^a {\chi_{\bar{Q}}}' F_R(-q)
\right] \times {\hat O}( brown\, muck) \right\}  + h.c.
,
\end{eqnarray}
where the ``form factor''
\begin{equation}
F_R(q)=\exp\left[\frac{i}{2}\left(\frac{q^+ R^3}{v^+}-{\vec
R}_\perp{\vec q}_\perp\right)\right]
\label{eq:form}
\end{equation}
arises from acting with the (kinematic !) displacement
operator on the position of the heavy quark/antiquark (from $x^-=0$, ${\vec
x}_\perp = {\vec 0}_\perp$ to $x^-=\pm R^3/2v^+$, ${\vec x}_\perp =
\pm {\vec R}_\perp/2$)
and $q$ is the {\it net} momentum transferred to the brown muck.
This rule holds irrespective of the number of gluons involved in
this process. ${\hat O}( brown\, muck)$ contains all the creation/destruction
operators acting on the brown muck as well as
the counterterms and, when necessary, the counterterm functions.
Note that ${\hat O}( brown\, muck)$ is the same for one or two heavy sources.
\item There is a static potential between the two heavy quarks.
The canonical Hamiltonian yields
$P^-_{HH} = g^2\,{\chi_Q^\dagger}' T^a \chi_Q\,
\chi_{\bar{Q}}^\dagger T^a
{\chi_{\bar{Q}}}'\, \delta^{(2)}({\vec R}_\perp)|R^3|v^+$.
In general, there will be a more complicated dependence on ${\vec R}$
which has to be determined by demanding self-consistency.
For example, singularities arising from exchange of gluons
with low $q^+$ between the two sources should cancel (nonperturbatively)
with the IR behavior of the instantaneous potential in $P_{HH}^-$ [5].
\end{itemize}
Using Eq.(\ref{eq:deltae}), $V({\vec R})$ can thus be extracted as follows:
\begin{itemize}
\item[1.] For a given ${\vec R}$ and $v^+$, write down the effective
LF-Hamiltonian for the heavy pair interacting with the brown muck
(including the formfactors Eq.(\ref{eq:form}) and including all
the counterterms and counterterm functions which would also appear
in a ``heavy-light'' system).
\item[2.] The lowest eigenvalue $\delta E^{(1)}$ from Eq.(\ref{eq:deltae}),
i.e. the QCD-ground state in the presence of the two heavy sources,
is then equal to $V({\vec R})$.
\end{itemize}
The heavy quark potential thus calculated is equivalent to
the potential which a lattice theorist would extract from
an asymmetric rectangular Wilson loop. Since there is plenty
of ``quenched'' lattice data around, it would make sense to
omit light quarks completely in a first approach and to
focus on the pure glue part of the brown muck. However, the
formalism described above is so general that one could also
use it in a LF calculation that includes (dynamical) light quarks.
\vskip.3in
{\bf 5. Summary}
\vskip.1in
I have set up a very general scheme for computing the potential
energy between two heavy quarks from a LF calculation. The
scheme has been deliberately left very general in order to make
it easy for people using very different approaches to LF-QCD
to adopt the scheme and to calculate this important observable.
Calculating the heavy quark potential on the LF is useful for
testing rotational invariance and for comparing to lattice data and
fits to charmonium spectra. Compared to other LF calculations,
the main advantage is that the calculation can be done in the
pure glue sector (the two sources are static!) and one thus
does not have to bother about issues such as kinetic versus vertex masses and
zero modes of the quark.
Of course, if the effective $Q\overline{Q}$ potential violates
rotational invariance, this would perhaps also show up indirectly
in the charmonium spectrum. However, a direct calculation of the
$Q\overline{Q}$ potential, as I propose here, will reveal the
source of the violation more clearly: instead of
comparing charmonium states, which are polarized in various directions,
one can just orient the
``dumbbell'' in different directions.

Conceivable applications of this scheme are, for example:
\begin{itemize}
\item helping to pin down the counterterm functions in the LF
Hamiltonian: one could imagine parameterizing
the noncovariant counterterm functions in the
Hamiltonian (basing the ansatz on
some superposition of perturbation theory, power-counting
and intuition) by a number of ``coupling constants.'' Demanding rotational
invariance of the $Q\bar{Q}$ potential at all distances provides
constraints in this coupling constant space, thus complementing renormalization
group studies of LF Hamiltonians.
\item fixing the transverse scale and determining the
renormalization constants for the
various couplings which appear in the transverse
lattice formulation of LF QCD.
\end{itemize}
Another potential use of the above formalism derives from
a curious mathematical observation: usually, in a LF Hamiltonian,
all energies scale like $(\mbox{momentum})^{-1}$ and thus an unconstrained
variational calculation (total momentum not fixed) of the
energy leads to nonsense: infinite LF momentum, zero energy for
all states. This is quite cumbersone because this usually
implies that one has to build momentum conservation into
the variational ansatz --- thus making it often difficult to calculate
matrix elements of the interaction in the multiparticle (I mean more than
three) sectors of the wavefunction. \footnote{Unless one uses a plane
wave basis.} This is not the case for the
effective LF Hamiltonian for a heavy quark system (\ref{eq:deltae}),
where first of all the brown muck momentum is {\it not} conserved,
and second ``runaway solutions'' (momentum $\rightarrow \infty$)
are prevented due to the term proportional to the
brown muck momentum on the r.h.s. of Eq.(\ref{eq:deltae}),
which arises from expanding the heavy quark kinetic term and acts like
a Lagrangean multiplier. For heavy quark systems it is therefore
easy to write down an ansatz for a orthonormalized set of
manybody wavefunctions that still allows simple evaluation of
matrix elements. Many body techniques can thus more easily be applied
to heavy quark systems on the LF than to hadrons with finite mass on the LF.

\vskip.3in
{\bf References}
\vskip.1in

1. S.D. G{\l}azek and R.J. Perry, Phys. Rev. D{\bf 45}, 3734 (1992).

2. M. Burkardt, Phys. Rev. D{\bf 46}, 1924 (1992); 2751 (1992).

3. M. Burkardt and E.S. Swanson, Phys. Rev. D{\bf 46}, 5083 (1992).

4. See, e.g. K.G. Wilson's lectures, these proceedings.

5. R.J. Perry, Invited lectures presented at Hadrons 94, Gramado,
Brazil, April, 1994.
\end{document}